\documentclass[reprint,aps,nofootinbib,preprintnumbers,amsmath,amssymb,11pt]{revtex4}
\usepackage{graphicx}% Include figure files
\usepackage{dcolumn}% Align table columns on decimal point
\usepackage{bm}% bold math
\usepackage{natbib}
\newcommand{\fft}[2]{{\frac{#1}{#2}}}

\newcommand{\beq}{\begin{equation}}
\newcommand{\eq}{\end{equation}}
\newcommand{\bea}{\begin{eqnarray}\displaystyle}
\newcommand{\ea}{\end{eqnarray}}
\newcommand{\es}{\eta/s}

\newcommand{\nn}{\nonumber}

\newcommand{\etas}{\frac{\eta}{s}}
\def\calo         {{\cal O}}
\def\w{\omega}

\begin{document}

\preprint{DAMTP-2011-59}
\preprint{MIFPA-11-34}

\title{THE SHEAR VISCOSITY TO ENTROPY RATIO: \\A STATUS REPORT}

\author{Sera Cremonini$ ^{\,\clubsuit,\spadesuit}\,$}
\email{S.Cremonini@damtp.cam.ac.uk}
\affiliation{$ ^\clubsuit$ Centre for Theoretical Cosmology, DAMTP, CMS,\\
\it University of Cambridge, Wilberforce Road, Cambridge, CB3 0WA, UK \\
%[.5em]
\it $ ^\spadesuit$ George and Cynthia Mitchell Institute for Fundamental Physics and Astronomy\\
\it Texas A\&M University, College Station, TX 77843--4242, USA}

\date{\today}

\begin{abstract}
This review highlights some of the lessons that the holographic gauge/gravity duality has taught us regarding the
behavior of the shear viscosity to entropy density in strongly coupled field theories.
The viscosity to entropy ratio has been shown to take on a very simple \emph{universal} value in all gauge theories with
an Einstein gravity dual.
Here we describe the origin of this universal ratio, and focus on how it is modified by
generic higher derivative corrections corresponding to curvature corrections on the gravity side of the duality.
In particular, certain curvature corrections are known to push the viscosity to entropy ratio below its universal
value. This disproves a longstanding conjecture that such a universal value represents a strict lower bound for any fluid in nature.
We discuss the main developments that have led to insight into the violation of this bound,
and consider whether the consistency of the theory is responsible for setting a fundamental lower bound on
the viscosity to entropy ratio.
\end{abstract}

\maketitle
\newpage

\tableofcontents

%\keywords{Keyword1; keyword2; keyword3.}
\newpage
\section{Introduction}
\label{Intro}

The development of the AdS/CFT correspondence \cite{Maldacena:1997re,Gubser:1998bc,Witten:1998qj}, or the gauge/gravity
duality more broadly, has led to new insights into features of quantum gravity and black hole physics, but also
-- and perhaps more surprisingly -- to a new paradigm for probing strongly coupled gauge theories.
At the core of the AdS/CFT correspondence is a relation between string theory in certain curved backgrounds (asymptotic to anti-de Sitter space)
and certain quantum field theories (conformal field theories) in one lower dimension -- hence, the correspondence is holographic.
Moreover, it is a strong/weak coupling duality -- when the gauge coupling is large, the field theory is in a
non-perturbative regime, whereas the string theory side can be approximated by its effective low-energy
description, classical supergravity. Precisely because of this feature, the duality allows us to describe
strongly interacting gauge theories in terms of weakly coupled gravitational systems.

Although the original and best studied example of the AdS/CFT correspondence involves theories with a large amount of supersymmetry,
the duality has been extended to a variety of other cases, and has taken on a life of its own.
In addition to top-down studies of AdS/CFT, based on string/M-theory constructions,
there have been a number of phenomenological bottom-up approaches, with applications ranging from QCD-like theories to condensed matter systems.
In particular, the duality has proven to be a useful tool for probing thermal as well as hydrodynamic properties
of field theories at strong coupling.
The fact that it can be used to understand dynamics is
especially valuable, given that there are very few theoretical tools available for doing so --
while lattice methods work well for thermodynamic quantities,
they are much less successful in describing real-time processes.

Obtaining a `microscopic' description of strongly correlated systems is extremely challenging.
However, one is typically interested in their macroscopic behavior
-- at large distances and long time scales -- and
in this regime, a system generically exhibits features which are \emph{universal} -- independent of the fine
details of the underlying microscopic description.
This explains the  importance, within the AdS/CFT program,
of identifying quantities which exibit universal behavior,
as they will allow us to gain intuition about real systems that lie in the same universality class.

\subsection{The quark gluon plasma and the shear viscosity to entropy ratio}

A quantity which has played a prominent role in AdS/CFT applications to the realm of strongly coupled thermal
gauge theories is the ratio of shear viscosity $\eta$ to entropy density $s$.
The interest in this ratio  was sparked by the fact that it was shown to take on a very simple value \cite{pss,Buchel:2003tz}
\beq
\frac{\eta}{s}=\frac{\hbar}{4\pi k_B}
\eq
in all gauge theories with Einstein gravity duals, and is therefore \emph{universal} in the sense discussed above.
Furthermore, elliptic flow measurements from the relativistic heavy ion collider (RHIC)
indicate that this ratio is also unusually small for the strongly coupled QCD quark gluon plasma and even appears
to yield roughly $\eta/s \sim 1/4 \pi$ (a recent estimate \cite{Song:2010mg} for the viscosity to entropy ratio
yields $\etas \leq 2.5 \frac{1}{4\pi}$).
We refer the reader to \cite{Teaney:2003kp,Luzum:2008cw} for some early references on the
RHIC results and the range of $\es$, and to \cite{Luzum:2010ag,Nagle:2011uz,Shen:2011eg}
for more recent ones including discussions of the first LHC results.

The universality of the simple result\footnote{From now on, we will set $\hbar$ and $k_B$ equal to one.}
$\eta/s=1/4\pi$ and its order-of-magnitude agreement with the RHIC data
have driven a large effort to apply holographic methods to the calculation of various transport coefficients of the QCD guark gluon plasma
(see \cite{CasalderreySolana:2011us} for a recent review, and \cite{Sinha:2009ev} for a discussion focusing on the shear viscosity).
The gauge theories which are most amenable to such holographic studies are somewhat exotic compared to QCD.
However, for temperatures that are not too much above that of the deconfinement transition,
the quark gluon plasma shares many of the properties of the strongly coupled $\mathcal N=4$ SYM plasma.
As long as one is not too close to the deconfinement transition -- where the bulk viscosity, which signals deviation from
conformality, is expected to be large -- the plasma is nearly conformal.
Moreover, it is strongly coupled, as reflected in the very small ratio of shear viscosity
to entropy density, indicating that the quark gluon plasma behaves as a nearly ideal fluid \cite{Adcox:2004mh}.
This is in sharp contrast with weak coupling calculations in thermal field theory, which predict a very large ratio:
\beq
\etas \sim \frac{1}{\lambda^4 \log(1/\lambda^2)} \gg 1 \, .
\eq
Finally, some of the properties of the plasma may be universal -- as is the case for $\es$ --
and therefore insensitive to the fine details of a particular construction.

The universal result $\es=1/4\pi$ holds in a remarkably wide class of theories and situations, e.g.
with various gauge groups and matter content, with or without chemical potentials, with non-commutative spatial directions or in
external background fields. It is also well understood that higher curvature corrections
in the dual gravitational theory modify this ratio. In fact, it was shown that for certain
theories such corrections produce even lower values of $\es$, thus disproving a longstanding
conjecture -- the celebrated KSS bound \cite{Kovtun:2003wp,u2} -- that $1/4 \pi$ represented a \emph{strict lower bound} for the viscosity of any
fluid in nature:
\beq
\label{KSSbd}
\frac{\eta}{s} \geq\frac1{4\pi} \, .
\eq

The amount by which the bound is violated is controlled by the couplings of certain higher curvature corrections.
On the gauge theory side of the duality, the corrections which have been shown to violate the KSS bound
correspond to finite $N$ effects which, in turn, can be parametrized in terms of the central charges of the conformal
field theory under consideration. On the other hand, the corrections which respect the bound correspond to finite $\lambda$
effects.
Thus, while the KSS bound does not hold in general, it is satisfied in the limit of infinite number of colors, $N \rightarrow \infty$.
Even though it is now clear that (\ref{KSSbd}) is violated by $1/N$ corrections, the question of the existence of a possible
(new) lower bound for $\es$ is still open.
Furthermore, independently of this issue, holographic calculations in the presence of higher curvature interactions are valuable for
getting a better handle on the dependence of $\es$ on the various physical parameters in the theory, as well as for
broadening the universality class of conformal gauge theories under study.

This review will highlight some of the lessons we have learned in the context of the gauge/gravity duality about the
behavior of the shear viscosity to entropy ratio in strongly coupled field theories.
In Section \ref{Univ} we introduce the notion of viscosity and review some
of the properties of $\es$ in theories whose gravitational description is Einstein gravity.
In Section \ref{HDsection}, after stating the KSS bound, we focus on the
structure of the corrections to the leading $\es=1/4\pi$ result in the presence of
higher derivative (curvature) terms.
We consider both top-down string theory constructions as well as more phenomenological bottom-up approaches,
and show that the bound is violated generically by curvature-squared corrections to the leading supergravity approximation.
Section \ref{Consist} discusses issues related to the consistency of the relativistic quantum field theory describing the plasma,
in the presence of higher-derivative terms.
In particular, it focuses on the question of whether microscopic constraints are generically responsible for setting a lower
bound on $\es$, and describes an example which suggests that this is not the case.
Although the review attempts to touch on the main developments in the field, several topics will be omitted for the sake of brevity.

%%%%%%%%%%
%%%%%%%%%%
\section{The universality of $\eta/s$}
\label{Univ}
%%%%%%%%%%
%%%%%%%%%%

\subsection{Viscosity preliminaries}

We start with a brief review of how the notion of viscosity arises in the context of finite temperature field theory.
For a plasma slightly out of equilibrium, the stress-energy tensor in the local rest frame,
in which the three-momentum density $T_{0i}$ vanishes, is given by the constitutive relation\footnote{The
hydrodynamic stress tensor is an expansion
in powers of derivatives of the fluid velocity. Here we are stopping at linear order in derivatives.}:
\beq
T_{ij} = p \, \delta_{ij} -\eta \left(\partial_i u_j +\partial_j u_i - \frac{2}{3} \delta_{ij} \partial_k u_k \right)
- \zeta \, \delta_{ij} \, \partial_k u_k \, .
\eq
Here $u_i$ is the fluid velocity, $p$ the pressure and $\eta$ and $\zeta$ denote, respectively, the shear and bulk viscosities.
If the theory is conformal the stress-energy tensor is traceless,
the energy density and pressure obey
$\rho=T^0_0=3p$ and the bulk viscosity -- being proportional to the trace of the stress tensor -- vanishes by construction, $\zeta=0$.

Among the various methods developed for computing the shear viscosity from supergravity \cite{pss,Kovtun:2003wp,Policastro:2002se},
the most straightforward one is based on the celebrated Kubo formula, which relates the transport coefficients
of a slightly non-equilibrium system to real-time correlation functions, computed in an equilibrium thermal ensemble.
For the shear viscosity, the relevant correlator is that of the \emph{shear} component of the stress tensor,
\beq
\label{etaFT}
\eta = \lim_{\omega\rightarrow 0} \, \frac{1}{2\omega}\int dt d\vec{x} \, e^{i\omega t} \langle [T_{xy}(t,\vec{x}),T_{xy}(0,0)]\rangle \, ,
\eq
which in turn can be expressed in terms of the retarded Green's function:
\beq
\label{etaFT2}
\eta =  - \lim_{\omega\rightarrow 0} \, \frac{1}{\omega} \; \text{Im}\, G^R_{xy,xy} (\omega,0) \, .
\eq
The relation above holds for generic transport coefficients:
\beq
\label{generictransport}
\chi = -\lim_{\omega\rightarrow 0} \, \lim_{\vec{k}\rightarrow 0}\, \frac{1}{\omega} \, \text{Im}\, G^R (\omega,\vec{k})\, .
\eq
The small frequency and zero momentum limits in (\ref{generictransport}) are nothing but a reflection of the fact that we are in the large-distance,
long-time scale regime. In other words, the transport coefficients are just parameters in an
\emph{effective low-energy description} -- such as hydrodynamics --
and, once specified, determine the macroscopic behavior of the medium.

In the gauge/gravity duality dictionary, the object with is dual to the field theory stress-energy tensor is the metric. Thus,
on the gravity side of the correspondence, the shear viscosity can be computed by adding small (shear) metric fluctuations
$\phi=h^x_y$ to the appropriate black brane metric.
From the effective action for $\phi$ one can then derive the retarded two-point function for the $T_{xy}$ component of the boundary CFT stress tensor, and read off $\eta$ in the small $k$ and $\omega$ limits.

\subsection{$\frac{\eta}{s}=\frac{1}{4\pi}$ and the graviton absorption cross-section}

Black brane solutions of type II supergravity and their near-horizon geometry
are central elements of the AdS/CFT correspondence.
In particular, the low-energy description on the world-volume of $N$ $D3$-branes is that of ${\cal N}=4$ $U(N)$ supersymmetric Yang-Mills (SYM) theory.
When the number of $D3$-branes and the 't Hooft coupling are large, $N\rightarrow \infty$ and $\lambda=g_{YM}^2 N\to \infty$ respectively,
the curvature of the brane geometry is small enough that classical supergravity becomes a good description of the system.
In this regime, the appropriate geometry is that of $AdS_5 \times S^5$, the near-horizon geometry of the stack of $D$-branes.
Thus, we see the emergence of two different descriptions of the same physics -- one in terms of a strongly coupled gauge theory
living on the branes, and the other in terms of classical supergravity on an appropriate background.

Working in the framework of the gauge theory/gravity correspondence,
Policastro, Son and Starinets \cite{pss} computed the ratio of shear viscosity to entropy density for the
${\cal N}=4$ $SU(N)$ SYM plasma,
in the planar limit and for infinitely large 't Hooft coupling,
finding:
\begin{equation}
\frac{\eta}{s}=\frac{1}{4\pi} \, .
\label{univ}
\end{equation}
This remarkably elegant result is a consequence of the fact that the right-hand side of the Kubo formula (\ref{etaFT})
is proportional to the classical absorption cross section of gravitons by black three-branes \cite{Klebanov:1997kc,Gubser:1997yh}.

More precisely, the absorption cross section $\sigma$ of a graviton incident on a brane with energy $\omega$,
and polarized parallel to the brane (e.g. along the xy directions),
is related to the field theory stress-tensor correlator via
\beq
\label{sigma}
\sigma = \frac{\kappa^2}{\omega}\int dt d\vec{x} e^{i\omega t} \langle [T_{xy}(t,\vec{x}),T_{xy}(0,0)]\rangle \, .
\eq
Here $\kappa=\sqrt{8\pi}G$, where $G$ is the ten-dimensional gravitational constant.
Comparison with (\ref{etaFT}) shows that in the low-frequency limit the cross section can be related directly to the shear viscosity:
\beq
\eta=\frac{1}{2\kappa^2} \, \sigma(\omega=0) \, .
\eq
The simple relation (\ref{univ}) follows in a straightforward way once we recall that the
absorption cross section
is equal to the black-brane horizon area\footnote{Here $r_0$ denotes the horizon radius, and we are setting the $AdS$ radius equal to one.}
\beq
\sigma(0)=\pi^3 r_0^3 \, ,
\eq
which in turn is proportional to the entropy.
Written in terms of the temperature of the system $T=r_0/\pi$ and the number of D-branes $N=2\pi^{5/2}/\kappa$,
the shear viscosity becomes \cite{pss}:
\beq
\eta = \frac{\pi}{8} N^2 T^3 \, .
\eq
The result (\ref{univ}) is immediately reproduced once we combine this expression with the well-known form of
the entropy \cite{Gubser:1996de}
\beq
s = \frac{\pi^2}{2} N^2 T^3 \, .
\eq

It was later argued in \cite{u1} that (\ref{univ}) is in fact a \emph{universal} result\footnote{An exception is the case of
anisotropic fluids. A way to obtain deviation form this universality by breaking the rotational symmetry spontaneously was shown in
\cite{Natsuume:2010ky,Erdmenger:2010xm}.} in all gauge theory plasma
at infinite coupling holographically dual to classical, two-derivative (Einstein) gravity --
regardless of the matter content, the amount
of supersymmetry and the presence of conformality.
More recently, (\ref{univ}) has been shown to hold even for certain extremal black holes, i.e. at zero temperature \cite{Edalati:2009bi}.
Thus, for any theory describing Einstein gravity,
${\cal L} = R -\Lambda - F^2 + \ldots$,
one is guaranteed to find the result $\eta/s = 1/4\pi$.
Further generalizations and proofs of the shear viscosity universality theorem appeared in \cite{u3,bu,u5,u6,u7}.

\section{The conjectured shear viscosity bound and higher derivative corrections}
\label{HDsection}

Remarkably, the holographic result (\ref{univ}) is in agreement with a simple quasi-particle
picture of hydrodynamic transport. In fact, a naive dilute gas estimate gives a relation for the shear viscosity to entropy ratio
in terms of the average particle momentum $p$ and the mean free path $\ell_{mfp}$,
\beq
\frac{\eta}{s} \sim p \, \ell_{mfp}  \, ,
\eq
which seems to suggest a quantum mechanical bound \cite{ss} of the form
\begin{equation}
\frac{\eta}{s}\gtrsim {\cal O}\left(1\right) \, .
\label{bound0}
\end{equation}
This fact, along with the universality of $\eta/s = 1/4\pi$ in Einstein gravity and
the observation that all known fluids seem to have larger viscosity
to entropy density ratios\footnote{This includes superfluid helium, for which $\eta/s \sim 8 \, \frac{1}{4\pi}$,
as well as trapped $^6$Li at strong coupling \cite{Rupak:2007vp,Son:2007vk}.},
led Kovtun, Son and Starinets (KSS) to conjecture a bound for {\it any fluid} in nature \cite{Kovtun:2003wp,u2}:
\begin{equation}
\frac{\eta}{s}\ge \frac{1}{4\pi} \, .
\label{bound}
\end{equation}
%

%%%%
\subsection{Higher derivative corrections to the shear viscosity bound}
%%%%

A natural way to test the validity of the KSS conjecture is to ask how the viscosity to entropy ratio
is affected by the presence of higher derivative corrections.
On the gravity side of the correspondence, these are curvature corrections of the schematic form\footnote{Here $R^n$ denotes schematically all contractions involving $n$ Riemann tensors.}
\beq
{\cal L} = R-\Lambda -F^2 + \alpha^\prime R^2 + \alpha^{\prime \,2} R^3 + \alpha^{\prime \,3} R^4 + \ldots
\eq
while on the gauge theory side they correspond to finite $\lambda$ and finite $N$ effects.
Although higher derivatives theories lead generically
to undesirable features (a modified graviton propagator, ill-posed Cauchy problem, no generalization of the Gibbons-Hawking boundary term),
such pathologies are only problematic at the Planck scale,
and are therefore absent if the couplings of the higher derivative terms are perturbative\footnote{For a generalization of the Gibbons-Hawking
boundary term in theories with generic curvature-squared corrections, in the presence of R-charge, see \cite{Cremonini:2009ih}.},
as is the case in string theory reductions.
In fact, incorporating curvature corrections is natural from an effective field theory point of view -- after all, supergravity is
just the low-energy description of string theory.
Corrections are also of interest from a ``phenomenological'' point of view,
as they may bring observable quantities closer to their measured values.
For instance, this is particularly valuable for applications to the realm of the QCD quark gluon plasma,
for which the value of $\eta/s$ is still not known precisely, and seems to differ slightly from $1/4\pi$, as discussed in
Section \ref{Intro}.

The first test confirming the KSS bound came from considering $\mathcal N = 4$ SYM at large -- but finite --  't Hooft coupling \cite{Buchel:2004di}.
For further analysis and generalizations we refer the reader to \cite{f1,f2,Myers:2008yi,f4,Garousi:2008ai,Ghodsi:2009hg}.
In Type IIB supergravity on $AdS_5 \times S^5$, the gravitational theory dual to
$\mathcal N = 4$ SYM theory, the \emph{leading} corrections
coming from string theory are quartic in the curvature, schematically of the form $\sim \alpha'^3 \, R^4$.
Since in the AdS/CFT dictionary $\alpha'$ corrections to classical general relativity map to 't Hooft coupling corrections
to the field theory,
\beq
\alpha' = \frac{1}{\sqrt{g_{YM}^2 N}} = \frac{1}{\sqrt{\lambda}} \, ,
\eq
the terms quartic in the curvature yield \cite{Buchel:2004di,Buchel:2008sh} finite coupling corrections to $\es$ :
\beq
\frac{\eta}{s} = \frac{1}{4\pi}\Bigl[ 1+ 15 \, \zeta(3) \lambda^{-3/2} \Bigr] \, .
\eq
Thus, finite $\lambda$ corrections increase the shear viscosity to entropy ratio in the
direction consistent with the bound.
%\footnote{Further work on
%higher derivative corrections corresponding to finite $\lambda$ effects appeared in \cite{Garousi:2008ai,Ghodsi:2009hg},
%where the bound was shown to be satisfied in the regime of validity of the supergravity approximation.}.
The effects of higher derivative terms involving five-form flux -- in the context of finite $\lambda$ corrections to the leading
supergravity approximation -- were considered for the first time in \cite{Myers:2008yi}.

A phenomenological counterexample to the KSS bound  was constructed in \cite{Cohen:2007qr}, where it was shown that (\ref{bound})
could be violated by increasing the number of species in the fluid while keeping the dynamics essentially independent of the species type.
Unfortunately, this particular example does not have a well-defined relativistic quantum field theory completion \cite{Son:2007xw}.
A violation of the bound was also observed by \cite{Brigante:2007nu,Brigante:2008gz} in \emph{effective} theories of higher derivative gravity.
More precisely, \cite{Brigante:2007nu} considered a special combination of curvature-squared corrections,
packaged in the well-known Gauss-Bonnet term,
\beq
S =\frac{1}{16 \pi G_N }\int d^5x \sqrt{-g}\left[ R - 2\Lambda +
\frac{\lambda_{GB}}{2}\ L^2 \left(R^2-4 R_{\mu\nu}R^{\mu\nu}+R_{\mu\nu\rho\lambda}R^{\mu\nu\rho\lambda}\right) \right],
\label{Brigante}
\eq
and found a contribution to the shear viscosity to entropy ratio of the form
\beq
\frac{\eta}{s}=\frac{1}{4\pi}\left[ 1-4 \lambda_{GB} \right] \, ,
\eq
implying violation of the bound for $\lambda_{GB}$ positive.
However, one should keep in mind that the main question was whether the bound was satisfied in quantum field theories that
allowed for a  \emph{consistent ultraviolet completion}.

In \cite{Kats:2007mq} Kats and Petrov put forth the first consistent example of a relativistic
quantum field theory which violates the KSS viscosity bound
--- the four-dimensional ${\cal N}=2$ $Sp\,(N)$ superconformal gauge theory plasma coupled to four
hypermultiplets in the fundamental representation
and one hypermultiplet in the antisymmetric representation.
The gravitational theory dual to this setup is Type IIB string theory on $AdS_5 \times S^5/\mathbb{Z}_2$,
which can be viewed as the decoupling limit of N D3-branes overlapping with
a collection of $D7$-branes and an orientifold $7$-plane.
Unlike for $AdS_5 \times S^5$,  in this theory curvature corrections start at quadratic order.
While the analysis of \cite{Kats:2007mq} was done in D spacetime dimensions, here we review only the five-dimensional case.
For generic curvature-squared corrections of the form
\beq
S= \frac{1}{16\pi G_N} \int d^5x \sqrt{-g} \left(R-2\Lambda+\alpha_1 R^2 + \alpha_2 R_{\mu\nu}R^{\mu\nu}
+\alpha_3 R_{\mu\nu\rho\lambda} R^{\mu\nu\rho\lambda} \right) \, ,
\eq
the viscosity to entropy ratio takes the form \cite{Kats:2007mq} \footnote{The fact that the result
(\ref{katspetrov}) is independent of the $\alpha_1$ and $\alpha_2$ terms is expected, since the $R^2$ and $R_{\mu\nu}R^{\mu\nu}$
terms can be eliminated through a field redefinition.}
\beq
\label{katspetrov}
\frac{\eta}{s}=\frac{1}{4\pi}\left[ 1- 8 \, \frac{\alpha_3}{L^2}\right] \, ,
\eq
where $L$ is the AdS radius, which obeys $-2 \Lambda L^2=12$.
Thus, the shear viscosity bound is violated for $\alpha_3 >0$. In the set-up of \cite{Kats:2007mq} the $\alpha_3$ term
can be seen to come from the effective action on the worldvolume of the $D7$-branes/$O7$-plane system.
It is fixed in terms of the matter content of the dual CFT -- more precisely, it can be related to the central charges
of the dual CFT, via the holographic Weyl anomaly.

\subsection{The holographic Weyl anomaly}
\label{Weyl}

In fact, the couplings of the curvature corrections $\sim R^2$ -- which are geometric data -- can be related to the central charges of the dual CFT
by making use of the trace anomaly.
For a four-dimensional field theory in a curved background, the Weyl anomaly may be parametrized by two
coefficients, $a$ and $c$, the central charges,
\begin{equation}
\langle T^\mu{}_\mu\rangle_{\,CFT} =\frac{c}{16\pi^2} \, (\text{Weyl})^2 -\frac{a}{16\pi^2} \,(\text{Euler})\,,
\label{confag}
\end{equation}
where the four-dimensional Euler density and the square of the Weyl curvature are:
\begin{equation}
\label{ejdef}
\text{Euler}= R_{\mu\nu\rho\lambda}^2- 4 R_{\mu\nu}^2 + R^2 \, ,
\quad \quad (\text{Weyl})^2 = R_{\mu\nu\rho\lambda}^2-2  R_{\mu\nu}^2 +\frac{1}{3} R^2\,.
\end{equation}
At the two-derivative level, the holographic computation of the $\mathcal N=4$ SYM Weyl anomaly gives
$a=c=N^2/4$ \cite{Henningson:1998gx}.
Combining this with the standard $AdS/CFT$ dictionary $N^2 = \pi L^3/2G_5$
 allows us to write
\begin{equation}
a=c=\fft{\pi L^3}{8G_5},
\label{eq:acrel2d}
\end{equation}
which has the advantage of being completely general, independent of the particular gauge theory dual.

There is by now a well-known prescription for obtaining the holographic Weyl anomaly for higher
derivative gravity, which was worked out in \cite{Blau:1999vz,Nojiri:1999mh}, and
later extended in \cite{Fukuma:2001uf} for general curvature squared terms.
The result is that, for an action of the form
\begin{equation}
e^{-1} {\mathcal L} = \frac{1}{2\kappa^2} \left(R+12g^2
+ \alpha R^2 + \beta R_{\mu \nu}^2 + \gamma R_{\mu \nu \rho \sigma}^2+\cdots
\right)\;,
\end{equation}
%\begin{equation}
%e^{-1} {\mathcal L} = \frac{1}{2\kappa^2} \left(-R+12g^2
%+ \alpha R^2 + \beta R_{\mu \nu}^2 + \gamma R_{\mu \nu \rho \sigma}^2+\cdots
%\right)\;,
%\end{equation}
%
the holographic Weyl anomaly may be written as \cite{Fukuma:2001uf}
\begin{equation}
\langle T^\mu{}_\mu\rangle
=\frac{2L}{16\pi G_5} \left[ \Bigl(- \frac{L}{24} + \frac{5\alpha}{3}
+ \frac{\beta}{3} + \frac{\gamma}{3} \Bigr) R^2
+ \Bigl( \frac{L}{8} - 5 \alpha - \beta
- \frac{3 \gamma}{2} \Bigr) R_{\mu \nu}^2 + \frac{\gamma}{2}
R_{\mu \nu \rho \sigma}^2 \right],
\label{eq:adsweyl}
\end{equation}
where $L$ is related to $g$ (to linear order) by
\begin{equation}
g=\fft1L\left[1-\fft1{6L^2}(20\alpha+4\beta+2\gamma)\right].
\end{equation}
Comparison of (\ref{confag}) with (\ref{eq:adsweyl}) then gives the curvature-squared correction to (\ref{eq:acrel2d}):
\begin{eqnarray}
a&=&\fft{\pi L^3}{8G_5}\left[1-\fft4{L^2}(10\alpha+2\beta+\gamma)\right]\nn\\
c&=&\fft{\pi L^3}{8G_5}\left[1-\fft4{L^2}(10\alpha+2\beta-\gamma)\right].
\end{eqnarray}
These are the key relations in connecting the couplings of the four-derivative terms in the Lagrangian -- which are
geometrical quantities -- to gauge theory data.
Note in particular that the combination
\beq
\label{ca}
\frac{c-a}{a} = \frac{\gamma}{L^2}
\eq
is sensitive to the presence of the $R_{\mu \nu \rho \sigma}^2$ term in the action.

%%%%%%%
\subsection{Finite $\lambda$ vs. finite $N$ effects}
%%%%%%%

We can now reinterpret the results of \cite{Buchel:2004di} and \cite{Kats:2007mq} in terms of the central charges of the dual CFT.
The model \cite{Buchel:2004di}, along with generalizations \cite{f4,Buchel:2008sh}, describes a superconformal
gauge theory plasma --- a consistent relativistic quantum field theory --- with the same
anomaly coefficients (central charges) $c=a$ in the trace of the stress-energy tensor.
The absence of curvature-squared corrections in the $AdS_5 \times S^5$ theory can be explained from the equality between the central charges in ${\cal N}=4$ SYM: if $a=c$ exactly, (\ref{ca}) ensures that there will be no $R_{\mu \nu \rho \sigma}^2$ term in the action.
This feature is a reflection of the large amount of supersymmetry in the theory.

The construction of \cite{Kats:2007mq} has a lower amount of supersymmetry, allowing for the presence of curvature-squared terms in the action.
The $\alpha_3$ coupling is then related to the dual central charges in the following way:
\beq
\label{alpha3KP}
\frac{\alpha_3}{L^2 / \kappa} = \frac{1}{16} \, \frac{c-a}{c} \, .
\eq
The matter content of the conformal field theory studied in \cite{Kats:2007mq} forces the central charges to be
\beq
a=\frac{1}{24}\left(12 N^2 + 12 N -1\right) \, , \quad c=\frac{1}{24}\left(12 N^2 + 18 N -2\right) \, .
\eq
Combining (\ref{katspetrov}) and (\ref{alpha3KP}) makes manifest the violation of the bound
\beq
\frac{\eta}{s}=\frac{1}{4\pi}\left[ 1- \frac{1}{2N}\right] + {\cal O}\left(\frac{1}{N^2}\right) \, ,
\eq
showing explicitly that, unlike \cite{Buchel:2004di}, these corrections correspond to finite $N$ effects.

What we have seen is that the violation of the KSS bound can be traced back to the inequality of the central charges of the dual CFT,
assuming that $c-a>0$. The latter condition was shown to be generic in superconformal field theories with unequal central charges
in \cite{Buchel:2008vz}, where a large class of examples of CFTs violating the bound were considered.
Also, for holographic models of RG flow and versions of the c-theorem in the context of higher curvature gravity
we refer the reader to \cite{Myers:2010tj,Myers:2010xs,deBoer:2011wk}.

%%%%%%%%
%%%%%%%%
\subsection{The role of finite chemical potential}
%%%%%%%%
%%%%%%%%

There has also been significant interest in exploring how a non-vanishing chemical potential $\mu$ affects the hydrodynamics
of strongly coupled gauge theories.
In the gauge/gravity duality context, the chemical potential that is turned on is typically due to the presence of R-charge, and is not
the same as the more physically relevant chemical potential related to non-zero baryon number density.
Nonetheless, from a phenomenological viewpoint it is useful to be able to tune the R-charged chemical potential $\mu$
to mimic the effect of baryonic chemical potential, and ideally come closer to matching observations.
However, a caveat for applications to the quark gluon plasma is that the baryonic chemical potential $\mu_B$ relevant for RHIC is small,
$\mu_B/T \lesssim 0.15$, rendering the potential effects of a chemical potential rather limited.

In theories with an Einstein gravity dual, the chemical potential doesn't affect $\es$
\cite{Mas:2006dy,Son:2006em,Saremi:2006ep,Maeda:2006by}, as expected
from universality arguments,  even though the viscosity and entropy density independently depend rather non-trivially on $\mu$.
Thus, any modification to $\es$ in the presence of chemical potential can only come from higher derivative interactions in the gravitational theory.
In particular, it is interesting to determine whether the violation of the KSS bound gets larger or smaller as a function of chemical potential;
in fact, a large enough chemical potential could in principle lead to a restoration of the bound.

The effects of curvature-squared corrections on $\es$ at finite (R-charged) chemical potential
were explored in \cite{Myers:2009ij,Cremonini:2009sy}.
Motivated by the question of whether string theory would place interesting restrictions on the form
of $\es$, the focus of \cite{Cremonini:2009sy} was on \emph{supersymmetric} higher derivative terms, whose structure is highly
constrained\footnote{Four-derivative corrections
in the presence of a chemical potential have been partially discussed in \cite{Ge:2008ni,Cai:2008ph},
where $R^2$ and $F^4$ corrections were considered, respectively.
The supersymmetric Lagrangian, however, has $RF^2$ and $\nabla F \nabla F$-type
terms as well which were not previously considered.}.
The authors of \cite{Myers:2009ij} considered the most general set of four-derivative corrections to gravity in five dimensions,
coupled to a $U(1)$ gauge field with charge $\sim q$ and a negative cosmological constant:
\bea
S &=& \frac{1}{16 \pi G_N}
\int d^5 x \sqrt{-g}
\biggl[ R - 2\Lambda - \frac{F^2}{4} +
\frac{\kappa}{3} \epsilon^{\lambda\mu\nu\rho\sigma} A_\lambda F_{\mu\nu} F_{\rho\sigma} +  c_1 R_{\mu\nu\rho\lambda} R^{\mu\nu\rho\lambda}\biggr. \nn \\
  &+&   c_2 R_{\mu\nu\rho\lambda} F^{\mu\nu}F^{\rho\lambda} + c_3 (F^2)^2 + c_4 F^4 +
c_5 \epsilon^{\lambda\mu\nu\rho\sigma} A_\lambda R_{\mu\nu\tau\upsilon} R_{\rho\sigma}^{\;\;\;\tau\upsilon}  \biggr] \,.
\label{MPS}
\ea
In this case, the shear viscosity to entropy ratio takes the form:
\beq
\label{etasgen}
\etas = \frac{1}{4 \pi} \left(1-8 c_1 + 4(c_1+6c_2)\frac{q^2}{r_0^6}\right) \, ,
\eq
where $r_0$ denotes the horizon radius, and the chemical potential is $\mu \propto q/r_0^2$.

On the other hand, \cite{Cremonini:2009sy} worked in the framework of five-dimensional ${\cal N} =2$ gauged supergravity,
the gravitational theory dual to ${\cal N} =1$ SYM.
In that theory the four-derivative corrections to the leading order action include a mixed gauge-gravitational Chern-Simons term
$A \wedge \text{Tr}\,(R\wedge R)$, whose supersymmetric completion\footnote{The authors of \cite{Hanaki:2006pj}, using the superconformal
formalism, derived the off-shell action for $D=5$, $\mathcal N =2$ gauged supergravity at the four-derivative level.
Its on-shell generalization for the case of minimal supergravity was worked out in \cite{Cremonini:2008tw}.}
was obtained in \cite{Hanaki:2006pj}.
Thus, while in (\ref{MPS})
the couplings $c_i$ of the four-derivative terms are arbitrary, in the setup of \cite{Cremonini:2009sy} they
are constrained (and related to each other) by supersymmetry.

The five-dimensional $\mathcal N=2$ gauged supergravity Lagrangian up to four-derivative order takes the form \cite{Cremonini:2008tw}
\bea
S &=& \frac{1}{16 \pi G_N}
\int d^5 x \sqrt{-g}
\biggl[ R - 2\Lambda - \frac{F^2}{4} + \frac{1}{12\sqrt{3}} \, \epsilon^{\lambda\mu\nu\rho\sigma} A_\lambda F_{\mu\nu} F_{\rho\sigma} + \biggr. \nn \\
&+& \biggl. \bar{c}_2 \left(\frac{1}{8} \,C_{\mu\nu\rho\lambda} C^{\mu\nu\rho\lambda} +
\frac{1}{16\sqrt{3}} \, \epsilon^{\lambda\mu\nu\rho\sigma} A_\lambda R_{\mu\nu\tau\upsilon} R_{\rho\sigma}^{\;\;\;\tau\upsilon}
+\ldots \right) \biggr]\, ,
\label{eq:n=2sub}
\ea
where we have only written out a few noteworthy terms\footnote{Since the conventions adopted in \cite{Cremonini:2008tw} are non-standard, and differ
from those used in this review, we note them here. The metric in \cite{Cremonini:2008tw} is taken to be mostly minus,
the Riemann tensor is given by $[\nabla_\mu,\nabla_\nu]v^\sigma=R_{\mu\nu\rho}{}^{\, \sigma} \, v^\rho $ and
the Ricci tensor by
$R_{\mu\nu}=R_{\mu\rho}{}^\rho{}_\nu$.}, and the ellipses denote all the four-derivative terms related to
$C_{\mu\nu\rho\sigma}^2$ and $A \wedge \text{Tr}\,(R\wedge R)$ by supersymmetry.
The shear viscosity to entropy ratio then takes the simple form \cite{Cremonini:2008tw} :
\begin{equation}
\label{etaovers}
\fft\eta{s} = \frac{1}{4\pi} \Bigl[1 - \bar c_2(1+Q) \Bigr] \, ,
\end{equation}
where $Q$ is essentially the R-charge of the solution. To leading order in $\bar c_2$,
the charge $Q$ is related to the chemical potential $\mu$ and
the horizon radius $r_0$ via $\mu = r_0 \sqrt{3Q(1+Q)}$.
As shown in Section \ref{Weyl}, the geometric coupling $\bar{c}_2$ in (\ref{eq:n=2sub})
as well as the couplings $c_1,c_2$ in (\ref{MPS}), which control the strength of the derivative interactions,
can be related to the central charges of the dual theory by using either the trace anomaly result (\ref{ca})
or the R-current anomaly.
For $\mathcal N=2$ gauged supergravity (\ref{eq:n=2sub}) one finds \cite{Cremonini:2008tw}
\beq
\bar{c}_2 = \frac{c-a}{a} \, ,
\eq
allowing us to express the viscosity to entropy ratio in the following way:
\beq
\label{etassusy}
\etas = \frac{1}{4\pi} \left[1-\frac{c-a}{a}(1+Q)\right] \, .
\eq

Several interesting features can be seen from the results (\ref{etasgen}) and (\ref{etassusy}).
First of all, the bound violation, which occurs provided $c-a>0$, is a finite $N$ effect, since it depends crucially on $(c-a)/a$.
Secondly, since $Q\geq 0$, turning on $R$-charge enhances the amount of violation
-- a large enough chemical potential will not restore the bound, but rather exacerbate it.
Even though the bound is clearly violated, the violation is always small.
In fact, taking into account the allowed range $0\leq Q \leq 2$ for the charge, which can be extracted from the range
of the chemical potential,
it can be shown \cite{Cremonini:2009sy} that the deviation of
$\es$ from its universal value is constrained to be small:
\begin{equation}
\fft1{4\pi}\left(1-3\fft{c-a}a\right)\le\fft\eta{s}\le
\fft1{4\pi}\left(1-\fft{c-a}a\right) \,.
\end{equation}
Another important point to notice is that the only terms that affect the shear viscosity to entropy ratio are those
that have an \emph{explicit} dependence on the Riemann tensor. Thus, the supersymmetric completion of the curvature-squared terms does not seem
to play a role in the correction to $\es$ (although supersymmetry does constrain the values of the
central charges $c$ and $a$ of the dual CFT).
The dependence of $\es$ on terms containing the Riemann tensor only is reminiscent of Wald's entropy formula for
black holes with higher derivatives.
This similarily has sparked attempts to obtain an expression for $\es$ analogous
to Wald's entropy formula (see \cite{Brustein:2008cg} for an early, incomplete attempt),
as well as work on extracting transport coefficients from effective horizon graviton couplings \cite{Banerjee:2009wg,Paulos:2009yk}.
Moreover, while the dependence of $\eta$ and $s$ individually on the
$R$-charge is quite complicated, the ratio $\eta/s$ is remarkably simple, which seems suggestive of some (not yet understood) form
of universality.

Independently of the issue of the bound violation, allowing for a chemical potential has advantages from a
phenomenological point of view.
If one is interested in matching $\es$ to that measured in realistic strongly coupled systems,
turning on more parameters -- such as a chemical potential -- allows for additional tuning, which in turn might lead to better
agreement with data.
Finally, it is an interesting fundamental question whether violations of the
bound can be related to any constraints on the dual gravitational side -- or
consistency requirements of the underlying string theory.
For instance, as suggested in \cite{Kats:2007mq},
there might be a connection between the signs of the higher derivative couplings required
to satisfy the weak gravity conjecture of \cite{ArkaniHamed:2006dz} and that violating the shear
viscosity to entropy bound. This avenue was explored in \cite{Cremonini:2009ih,Amsel:2010aj}.

%%%%%%
%%%%%%
\section{Causality violation and a lower bound for $\es$}
\label{Consist}
%%%%%%
%%%%%%

As was emphasized in the previous section, the violation of the KSS bound can be understood in terms of
the inequality of the central charges of the conformal field theory, $c\neq a$.
More precisely, the violation occurs when $c-a >0$.
Moreover, since $c-a \sim N$, this is a finite $N$ effect, and is not due to having a finite 't Hooft coupling.
The inequality of the central charges translates into particular higher-derivative
corrections to the supergravity approximation \cite{Blau:1999vz} which, to ensure reliable computations,
have to be regarded as being \emph{small}. As a result, the KSS bound violation
in holographic models realized in string theory is necessarily perturbative.

The work \cite{Kats:2007mq} convincingly established that the original KSS bound (\ref{bound}) could not be a quantitative formulation of a
loose quantum-mechanical bound (\ref{bound0}).
Still, the question remained as to whether or not a bound of the type (\ref{bound0}) existed\footnote{The systems
we are describing, however, behave like strongly coupled nearly ideal fluids,
so the inapplicability of (\ref{bound0}) should not be too surprising.}.
To probe how much $\es$ can be lowered below its universal value $1/4\pi$, one needs access to a model where the couplings
of the higher derivative terms are finite -- perturbatively small couplings are not strong enough to suppress $\es$ significantly.
Thus, we should emphasize that any \emph{finite} violation of the KSS bound will necessarily entail working in a
holographic \emph{model} of the gauge/gravity correspondence, rather than in a particular string theory realization.

\subsection{The Gauss-Bonnet plasma and causality violation}

A useful model for probing these questions is that of Gauss-Bonnet gravity \cite{Brigante:2007nu}:
\beq
S =\frac{1}{16 \pi G_N }\int d^5x \sqrt{-g}\left[ R - 2\Lambda + % \frac{12}{L^2} +
\frac{\lambda_{GB}}{2}\ L^2 \left(R^2-4 R_{\mu\nu}R^{\mu\nu}+R_{\mu\nu\rho\lambda}R^{\mu\nu\rho\lambda}\right) \right].
\label{gbg}
\eq
As we already mentioned in Section \ref{HDsection}, the shear viscosity to entropy ratio in this model
takes the form \cite{Brigante:2007nu,Brigante:2008gz}:
\beq
\frac{\eta}{s}=\frac{1}{4\pi}\left[ 1-4 \lambda_{GB} \right] \, .
\label{etasGB}
\eq
Gauss-Bonnet gravity is a special case of the well-known Lovelock theories, which have the nice feature of being theories of
two-derivatives in disguise.
As a consequence, black-hole solutions to (\ref{gbg}) are known for arbitrarily large values of the Gauss-Bonnet coupling
(however, to ensure a vacuum AdS solution and a boundary CFT dual \cite{Brigante:2007nu} the coupling should
satisfy $\lambda_{GB}\leq 1/4$).
A crucial observation is that, once $\lambda_{GB}$ is allowed to be large,
(\ref{etasGB}) seems to lead to an \emph{arbitrary}
violation of the KSS bound, even allowing for a vanishing viscosity to entropy ratio.
Precisely because of this feature, the Gauss-Bonnet plasma is a natural playground to explore how low $\es$ can become.

We should also note that, for $\lambda_{GB}\ll 1$, the Gauss-Bonnet model (\ref{gbg}) is equivalent, up to field redefinitions,
to the string theory example of Kats and Petrov \cite{Kats:2007mq}.
In terms of the Gauss-Bonnet coupling, the central charges read:
\begin{equation}
\frac{c-a}{c}=4\lambda_{GB}+{\cal O}\left(\lambda_{GB}^2\right) \, .
\label{gbi}
\end{equation}
Again, unlike the construction of \cite{Kats:2007mq}, the Gauss-Bonnet model (\ref{gbg})
is consistent  for arbitrary values of the coupling $\lambda_{GB} \leq 1/4$.
As such, it {\it defines} via the AdS/CFT correspondence a dual conformal gauge theory plasma,
with \emph{effective} central charges \cite{mam}
 \bea
c&=&\frac{\pi^2}{2^{3/2}}\,\frac{L^3}{\ell_P^3}\,
(1+\sqrt{1-4\lambda_{GB}})^{3/2}\,\sqrt{1-4\lambda_{GB}}\,,\nonumber\\
a&=&\frac{\pi^2}{2^{3/2}}\,\frac{L^3}{\ell_P^3}\,
(1+\sqrt{1-4\lambda_{GB}})^{3/2}\,\left(3\sqrt{1-4\lambda_{GB}}-2\right)\,,
 \label{central}
 \ea
and hence
 \beq
\frac{c-a}{c}=2\left(\frac{1}{\sqrt{1-4\lambda_{GB}}}-1\right)\,,
\label{defd}
 \eq
 which clearly matches (\ref{gbi}) when the coupling is perturbative.
Here we see a parallel with the construction of \cite{Cohen:2007qr}. One can identify a relativistic
quantum field theory as a holographic dual to (\ref{gbg}), with a shear viscosity
to entropy density ratio given by (\ref{etasGB}).
In turn, this leads to an arbitrary violation of the KSS bound
given appropriate choices of
the central charges of the theory.

To fully understand the apparent arbitrary violation of the KSS bound, one must address the question of the consistency of the
Gauss-Bonnet plasma
as a relativistic quantum field theory.
This was done in \cite{Brigante:2008gz,bm}. It was found that once
\begin{equation}
\lambda_{GB} > \frac{9}{100}\,,
\label{b1}
\end{equation}
the spectrum of excitations in the plasma contains modes that propagate
faster than the speed of light \cite{Brigante:2008gz}.
Likewise, for
\begin{equation}
\lambda_{GB} < -\frac{7}{36} \, ,
\label{b2}
\end{equation}
the plasma also contains microcausality violating excitations \cite{bm}.
Given (\ref{b1}) and (\ref{b2}) one is led to conclude that consistency
of the Gauss-Bonnet plasma as a relativistic QFT constrains its viscosity ratio to be\footnote{See \cite{Ge:2008ni,Ge:2009eh,Ge:2009ac}
for further studies.}
\begin{equation}
\frac{16}{25}\ \le\ 4\pi \frac{\eta}{s}\ \le\ \frac{16}{9} \, .
\label{visco}
\end{equation}
Exactly the same constraint arises by requiring ``positivity of energy''
measured by a detector in the plasma \cite{hof}.
Thus, the Gauss-Bonnet plasma is a concrete example in which a lower bound on $\es$ and consistency of the theory are correlated.

Additional studies exploring the connection between causality and positivity of energy bounds appeared in
\cite{deBoer:2009pn,Buchel:2009sk,mam,gc1,gc3,gc4,Camanho:2010ru}.
Gauss-Bonnet gravity in $AdS_7$ was analyzed in \cite{deBoer:2009pn}, where it
was found that $\etas \geq \frac{7}{16} \, \frac{1}{4\pi}$.
Constraints arising from the second-order truncated theory of hydrodynamics were analyzed in \cite{Buchel:2009sk}.
The case of quasi-topological gravity, a gravitational theory which includes curvature-cubed
interactions, was considered in \cite{mam}, where it was found that $\etas \gtrsim \left(0.41\right) \frac{1}{4\pi}$.
The authors of \cite{gc3,gc4,Camanho:2010ru} explored Lovelock gravities in arbitrary spacetime dimensions.
In particular, the results of \cite{Camanho:2010ru} suggest that $\eta/s$ can be made arbitrarily small by considering a Lovelock
theory of high enough order, which also implies increasing the number of spacetime dimensions.

%%%%%
\subsection{Transport properties vs. causality: IR vs UV}
%%%%%

The example of the Gauss-Bonnet plasma (and its generalizations discussed above)
appears to suggest a \emph{link} between the violation of a shear viscosity bound of the type (\ref{bound0}) and
the violation of microcausality/positivity of energy in the theory.
However, one should keep in mind that the shear viscosity is one of the couplings of
the effective hydrodynamic description of the theory at lowest momenta
and frequency,  \emph{i.e.} for:
\begin{equation}
{\rm IR:}\quad\quad\quad\quad {\w}\ll \min({T,\mu,\cdots})\,,\quad\quad\quad {|\vec{k}|}\ll \min({T,\mu,\cdots})\, .
\label{hydro}
\end{equation}
Here $\cdots$ stand for any microscopic scales of the plasma other than the temperature
and chemical potential(s) for the conserved charge(s).
On the contrary, the microcausality of the theory is determined by the propagation
of modes in exactly the opposite regime, \emph{i.e.} for:
\begin{equation}
{\rm UV:}\quad\quad\quad {\w}\gg \max({T,\mu,\cdots})\,,\quad\quad {|\vec{k}|}\gg \max({T,\mu,\cdots}) \, .
\label{causality}
\end{equation}
Thus, transport properties are determined by the IR features of the theory,
while causality is determined by the propagation of UV modes -- whose dynamics is not that of hydrodynamics.

A direct link between the features of the theory governing its microcausality
and its shear viscosity should only be possible if the {\it same phase} of the theory
extends over the entire range of energy scales --- from the infrared to the ultraviolet.
In other words, there must not be any phase transitions in the plasma.
This is certainly the case in the Gauss-Bonnet plasma -- since the plasma is conformal,
and temperature is the only available scale in the model, there can not be
any phase transition in the theory as a function of temperature.
The only free parameter of the model is the Gauss-Bonnet coupling constant $\lambda_{GB}$,
which determines both the shear viscosity ratio and its microcausality
properties. This explains the origin of the link between the two, observed in
\cite{Brigante:2008gz} and in its generalizations.

A holographic model which suggests that this link may not be generic was put forth in \cite{Buchel:2010wf},
and involves a generalization of the Gauss-Bonnet plasma realized in a theory
with a superfluid phase transition.
The model is based on the construction of \cite{Gubser:2009qm} describing a theory undergoing a
second order phase transition below a critical temperature $T_c\propto \mu$, associated with the spontaneous
breaking of a global $U(1)$ symmetry, and the generation of a condensate of an irrelevant operator $\calo_c$ :
\begin{equation}
\langle \calo_c\rangle \
\begin{cases}
=0\,,\qquad T>T_c\\
\ne 0\,,\qquad  T<T_c\,.
\end{cases}
\label{cond}
\end{equation}
The Lagrangian for the model constructed in \cite{Buchel:2010wf} is described by
\begin{equation}
{\cal L}=R-\frac{1}{3} F_{\mu\nu}F^{\mu\nu}+
\frac{2}{27}\,
\epsilon^{\lambda\mu\nu\sigma\rho}F_{\lambda\mu}F_{\nu\sigma}A_\rho
+{\cal L}_{scalar}+{\cal L}_{GB} \, ,
\label{ea1}
\end{equation}
where
\bea
{\cal L}_{scalar}&=&-\frac{1}{2}\left[\left(\partial_\mu\phi\right)^2+4 \phi^2 A_\mu A^\mu\right]+12+\frac{3}{2}\phi^2\,,
\label{lscalar1} \\
{\cal L}_{GB}&=&\beta \phi^4 \biggl(R^2-4 R_{\mu\nu}R^{\mu\nu}+R_{\mu\nu\rho\lambda}R^{\mu\nu\rho\lambda}\biggr) \, ,
\label{lgb}
\ea
and $\phi$ is the scalar field dual to the CFT operator $\calo_c$.
Thus, this model modifies the theory analyzed in \cite{Gubser:2009qm} by the addition of a (generalized) Gauss-Bonnet term.

While at high temperatures $T>T_c$ the background is that of an electrically charged $AdS$ black hole, for $T<T_c$
the black hole develops scalar hair.
The model of \cite{Buchel:2010wf} is engineered in such a way that the effective Gauss-Bonnet coupling -- being
proportional to the operator that condenses -- is only non-zero in the symmetry-broken phase,
so that:
\begin{equation}
\lambda_{GB}\bigg|^{effective}\
\begin{cases}
=0\,,\qquad {\rm UV}\\
\ne 0\,,\qquad  {\rm IR }\,.
\end{cases}
\label{luvir}
\end{equation}
For $T>T_c$ the gravitational theory is simply Einstein gravity, and therefore $\es=1/4\pi$.
On the other hand in the symmetry-broken phase $\es$ is corrected by the effective Gauss-Bonnet term\footnote{Note that
in the superfluid phase, \emph{i.e.} for $T<T_c$,  for the case of $\beta=0$ one still finds \cite{Buchel:2010wf} $\es=1/4\pi$,
as expected from universality.}. As the temperature is lowered and the coupling $\beta$ is varied,
$\es$ decreases well below the pure Gauss-Bonnet lower bound (\ref{visco}) found in \cite{Brigante:2007nu,Brigante:2008gz}
and does \emph{not} appear to have a lower bound\footnote{One should
keep in mind, however, that the temperatures probed in \cite{Buchel:2010wf}, although very low, do not reach $T=0$.}.
Furthermore, no superluminal causality-violating modes are detected, and the theory describing the plasma appears to be consistent\footnote{The
analysis of \cite{Buchel:2010wf} was performed entirely in the scalar channel, and instabilities could in principle come from other channels.
However, it should be noted that the causality violation responsible for the lower bound in \cite{Brigante:2007nu,Brigante:2008gz}
came precisely from scalar channel quasi-normal modes.}. In the analysis the full back-reaction of the scalar field is taken into account.

In this model, microcausality features -- which are governed by the unbroken phase -- are completely \emph{decoupled} from the physics
that determines the shear viscosity of the superfluid phase. Thus, the model \cite{Buchel:2010wf} appears to suggest that microscopic constraints, while
important for the general consistency of the plasma as a relativistic field theory, are not necessarily responsible for
setting the lower bound on $\es$.
This is just a reflection of the fact that hydrodynamic transport of a system is determined by its infrared properties,
which are not always connected to the microcausality of the theory.
It appears, then, that the question of a bound on $\es$ -- suggested by a quasi-particle picture of the fluid -- is still open,
and moreover that the physics that would determine it, if such a bound exists, is still not understood.

\subsection{Radial flow \emph{vs.} temperature flow}

An interesting property of the shear viscosity is that it behaves trivially under Wilsonian RG flow.
In fact, $\eta$ is related to the momentum $\Pi(r)$ conjugate to the shear metric fluctuation $\phi \equiv h_x^y$
in the following way:
\beq
\eta =  - \lim_{\omega\rightarrow 0} \, \frac{1}{\omega} \; \text{Im}\, G^R_{xy,xy} (\omega,0) =
\lim_{r,\omega\rightarrow 0} \frac{\Pi(r)}{i\omega \phi(r)} \, .
\eq
Schematically, the conjugate momentum can be extracted from the effective action for $\phi$ via:
\beq
\Pi (r) \equiv \frac{\delta S_\phi^{(2)}}{\delta \partial_r \phi} \, .
\eq
It can then be shown that the radial flow of $\Pi$ is trivial in the low-frequency (hydrodynamic) regime:
\beq
\label{trivialflow}
\partial_r \Pi = 0 + {\cal O} (\omega^2) \, .
\eq
Since in the gauge/gravity duality framework the radial coordinate of the geometry
corresponds to the energy scale of the dual field theory, (\ref{trivialflow})
indicates a trivial RG flow for $\eta$.

The relation (\ref{trivialflow}), which was originally established in the context of the membrane paradigm \cite{u6},
also holds in theories with curvature-squared corrections \cite{Myers:2009ij}, as has been verified explicitly in a number of calculations.
Furthermore, (\ref{trivialflow}) makes the computation of $\eta$ particularly simple: the conjugate momentum
can be computed at any radius.
In particular, it can be evaluated at the horizon, making the computation of $\eta$ rather straightforward.
Interestingly, although $\es$ is a near-horizon quantity, its behavior is not entirely trivial.
It is possible to have constructions in which the value of $\es$ in the IR differs from that in the UV, with a non-trivial
temperature flow connecting the two regimes.

The superfluid model of \cite{Buchel:2010wf} is precisely an example of such a behavior.
The value of the viscosity to entropy ratio above the superfluid transition, for $T>T_c$, is simply $\es=1/4\pi$.
However in the superfluid phase, for $T<T_c$, it is corrected by the presence of higher derivative interactions, and
is no longer just $1/4\pi$.
The precise value of $\es$ in this phase depends on the temperature of the system: the viscosity to entropy ratio
 flows as a function of $T/\mu$, from
$\es=1/4\pi$ for $T \sim T_c$ to values significantly lower than $1/4\pi$ as one probes lower and lower temperatures,
\emph{i.e.} goes deeper in the IR.
Thus, although $\eta/s$ does not flow in any Wilsonian sense,
in the construction of \cite{Buchel:2010wf} the decoupling of the physics in the UV from that of the IR is reflected
in the different behavior of $\eta/s$ in the two temperature ranges.
This decoupling is also reflected by the fact that the correction to $\es$ in the model of \cite{Buchel:2010wf} can no longer be expressed
in terms of the central charges of the UV CFT.

We expect that the type of ``decoupling'' observed in \cite{Buchel:2010wf} should be seen in other setups, not necessarily involving phase transitions.
For instance, a non-trivial scalar profile (allowing for a dilatonic field),
along with sufficiently different behavior of the gravitational background (\emph{e.g.} different
symmetries) in the IR vs. UV regions, should be enough to give rise to a temperature-flow for $\es$.
In particular, this type of setup would likely force $\es$ to take on one value in the high-temperature theory,
and a different one in the low-temperature theory, as in the construction of \cite{Buchel:2010wf}, but possibly without the need for a phase transition.

It might be interesting to ask whether the type of flow described above can be understood more systematically
in the framework of the recent attempts to refine notions of holographic RG flow \cite{Bredberg:2010ky,Nickel:2010pr,Heemskerk:2010hk,Faulkner:2010jy},
especially in the constructions of \cite{Nickel:2010pr,Faulkner:2010jy}.
This might lead to further insight not only into the structure of the shear viscosity, but also into the issue of the possible existence of a lower bound
for $\es$, whose physics is still not entirely understood.
Independently of $\es$, new valuable lessons might be learned by extending some of the analysis of \cite{Bredberg:2010ky,Nickel:2010pr,Heemskerk:2010hk,Faulkner:2010jy}
to the realm of fully-fledged viscous hydrodynamics.

%%%%%%%%
%%%%%%%%
%%%%%%%%
\section{Acknowledgements}
I would like to thank the Benasque Center for Science for hospitality during the final stages of this work.
This research has been supported by the Cambridge-Mitchell Collaboration in Theoretical
Cosmology, and the Mitchell Family Foundation.
%%%%%%%%
%%%%%%%%
%%%%%%%%

\end{document}